\documentclass[lettersize,journal,onecolumn]{IEEEtran}
\usepackage{algorithmicx}
\usepackage{algpseudocode}
\usepackage{algorithm}
\usepackage{array}
\usepackage[caption=false,font=normalsize,labelfont=sf,textfont=sf]{subfig}
\usepackage{textcomp}
\usepackage{url}
\usepackage{verbatim}
\usepackage{graphicx}
\usepackage{cite}
\usepackage{amsmath,amssymb,amsfonts}

\usepackage{subcaption}
\usepackage{xcolor}
\usepackage{caption}
\usepackage[normalem]{ulem}
\usepackage{makecell}
\usepackage{tikz}
\usetikzlibrary{arrows.meta,positioning,fit}

\begin{document}

\title{GNN-RSMA: An Interference Management Framework for a Large-Scale HAPS Network}
\author{Afsoon Alidadi Shamsabadi,~\IEEEmembership{Senior Member,~IEEE,} Animesh Yadav,~\IEEEmembership{Senior Member,~IEEE,} and~Halim~Yanikomeroglu,~\IEEEmembership{Fellow,~IEEE}
\thanks{Afsoon Alidadi Shamsabadi and Halim Yanikomeroglu are with the Carleton-NTN Lab at the Department of Systems and Computer Engineering, Carleton University, Ottawa, ON K1S 5B6, Canada (e-mail: afsoonalidadishamsa@sce.carleton.ca, halim@sce.carleton.ca). Animesh Yadav is with the School of Electrical Engineering and Computer Science, Ohio University, Athens, OH 45701 USA (e-mail: yadava@ohio.edu).}
}
\maketitle

\begin{abstract}
Integrating non-terrestrial networks (NTN) with terrestrial infrastructure is a key enabler of next-generation wireless systems, providing ubiquitous connectivity while meeting stringent rate and latency requirements. In particular, high altitude platform stations (HAPS) can complement terrestrial networks and jointly form vertical heterogeneous networks (vHetNets), extending coverage while delivering high-capacity, reliable, and low-latency connectivity for user equipments (UEs) including ground users and uncrewed aerial vehicles (UAVs). However, the high altitude deployment of HAPS establishes strong line-of-sight (LoS) links to UEs, creating highly correlated channels among UEs. Moreover, the wide coverage footprint of HAPS enables it to serve a large number of UEs, forcing limited radio resources to be shared among many UEs and resulting in significant intra-resource block (RB) interference.
To address this challenge, we propose an interference management scheme based on UE clustering and rate-splitting multiple access (RSMA). Specifically, the network is modeled as a heterogeneous graph, and a graph neural network (GNN) is developed to efficiently allocate the common and private RSMA powers, maximizing the minimum spectral efficiency (SE) in a fast and scalable manner. Simulation results demonstrate that the proposed GNN-RSMA interference management algorithm outperforms conventional multiple access schemes while achieving fairness and worst-user performance comparable to successive convex approximation (SCA)-based optimization at only a fraction of its computational cost.
\end{abstract}


\section{Introduction}\label{Introduction}
%%%%%%%%%%%%%%%%%%%%%%%%%%%%%%%%%%%%%%%%%%%%%%%%%%%%%%%%%%%%
High altitude platform stations (HAPS), operating at altitudes of $18$--$50$ km above the Earth's surface, offer significantly wider coverage than terrestrial base stations while providing lower latency and higher-capacity communications than satellites due to their closer proximity to the Earth. These advantages make HAPS a promising platform for achieving ubiquitous connectivity in next-generation wireless networks~\cite{6GVTM,HAPSSurvey}. Furthermore, HAPS can be integrated with terrestrial infrastructure to form vertical heterogeneous networks (vHetNets), enabling seamless connectivity for user equipments (UEs)~\cite{HAPSSurvey}. However, the propagation characteristics of HAPS differ significantly from those of terrestrial networks~\cite{IM_Magazine}. Specifically, the large coverage area and predominantly strong line-of-sight (LoS) links enable a HAPS to serve many UEs simultaneously. At the same time, as geographically separated UEs often appear closely spaced from the HAPS perspective, they tend to experience highly similar channel conditions. Consequently, resource sharing among UEs can lead to severe intra-tier interference, making effective interference management essential~\cite{GCPaper}.

HAPS are generally equipped with large antenna arrays, which can be leveraged to mitigate interference through spatial beamforming. However, the effectiveness of per-UE beamforming in HAPS systems is constrained by their high altitude and extensive coverage area. Due to the limited angular separability of many UEs from the HAPS perspective, distinguishing and serving individual UEs with dedicated beams becomes challenging. Moreover, designing beams and acquiring accurate channel state information (CSI) for a large number of UEs impose substantial computational and signaling overhead. To this end, we proposed a clustering-based beamforming framework in~\cite{GCPaper}, where UEs with similar angular characteristics are grouped into clusters and served by a common beam. This approach significantly reduces CSI acquisition and beamforming complexity while still exploiting the spatial degrees of freedom provided by the antenna array. In addition, orthogonal resource block (RB) allocation within each cluster eliminates intra-cluster interference.
However, as multiple clusters share the available communication resources, inter-cluster interference remains a critical challenge requiring effective interference management.

Various interference mitigation strategies across different network layers have been proposed for wireless networks~\cite{IM_5G,IM2,IM3}. Among these, advanced multiple access schemes, including rate-splitting multiple access (RSMA), have recently emerged as powerful schemes for interference management. RSMA provides flexible interference handling by enabling partial decoding and suppression of interference, thereby bridging the gap between treating interference as noise in spatial division multiple access (SDMA) and fully decoding interference in non-orthogonal multiple access (NOMA) schemes~\cite{Dizdar2020RSMA6G}. However, to achieve the full benefits of RSMA, the transmit powers of the common and private streams and the rates of the common streams must be optimized jointly.
In HAPS networks, the wide coverage footprint naturally supports a large number of UEs, leading to high-dimensional, non-convex resource allocation problems. Although successive convex approximation (SCA) can effectively obtain suboptimal solutions, its iterative nature incurs prohibitive computational complexity as the numbers of clusters and UEs increase~\cite{SCA}. This motivates artificial intelligence (AI)-driven approaches, such as deep neural networks (DNNs) and reinforcement learning (RL), that enable low-complexity, real-time resource allocation by directly learning mappings from network states to resource allocation decisions.
Among AI-driven approaches, graph neural networks (GNNs)~\cite{GNNSurvey} have emerged as promising solutions for large-scale wireless resource allocation problems due to their permutation invariance and ability to process graph-structured data. Unlike conventional DNNs and many RL-based approaches that rely on fixed-size state and action spaces, GNNs can process variable-sized graph representations. Consequently, they can seamlessly adapt to different numbers of UEs, clusters, and RBs, making them a natural fit for HAPS systems.

\subsection{Related Works}\label{RelatedWorks}
RSMA has attracted growing attention as a flexible multiple access scheme for interference management in multi-antenna and multi-carrier wireless communication systems~\cite{MMFRA_RSMA,Mehmet,Zhang,Huang}. In this context, the authors in~\cite{MMFRA_RSMA} derive a closed-form max-min fair (MMF) power and common-rate allocation solution for RSMA in a multi-antenna broadcast channel under a practical low-complexity beamforming design. Furthermore, the authors in~\cite{Mehmet} integrate RSMA with multi-numerology orthogonal frequency division multiplexing (OFDM) and formulate a weighted minimum mean square error (WMMSE)-based joint power and subcarrier allocation framework to mitigate multi-UE and inter-carrier interference. Beyond terrestrial and single-tier networks, the authors in~\cite{Zhang} develop a distributed deep-learning-assisted RSMA framework for interference management in space--air--ground integrated networks, achieving improved sum-rate performance with reduced computational cost. Similarly, the authors in~\cite{Huang} propose a robust RSMA-based secure precoding scheme for HAPS-assisted satellite networks that jointly optimizes satellite and relay transmissions under imperfect CSI to enhance secrecy performance.

Meanwhile, GNNs have emerged as powerful and scalable learning frameworks for resource allocation in wireless networks due to their ability to exploit graph structures and generalize across different network sizes. The authors in~\cite{GNN-ChannelPower} propose a GNN-based approach for joint channel and power allocation in heterogeneous wireless networks, while the authors in~\cite{MMF-TVT} develop a GNN-based MMF resource allocation framework for RSMA in a single-base-station scenario. Moreover, the authors in~\cite{Shen_GNN} formulate radio resource management as a graph optimization problem and demonstrate that message-passing GNNs generalize effectively to large-scale instances with high computational efficiency. Likewise, the authors in~\cite{Eisen_REGNN} introduce random-edge GNNs trained using an unsupervised primal--dual framework and establish their permutation equivariance and transferability across different network topologies. GNNs have also been integrated with optimization-inspired architectures, such as the WMMSE-unfolding GNN in~\cite{Chowdhury_WMMSE}, and extended to richer graph representations, including heterogeneous GNNs for multi-cell multi-user power allocation~\cite{Guo_HeteroGNN} and bipartite GNNs for scalable beamforming~\cite{Kim_Bipartite}. Recently, in \cite{GNN_vtc_2026}, the authors develop a GNN-based beamforming optimization framework for HAPS-assisted terrestrial networks, demonstrating improved energy efficiency and enhanced service for cell-edge UEs. A comprehensive survey of GNN applications in wireless communications is provided in~\cite{He_GNNSurvey}.

Despite recent advances, existing GNN-based resource allocation frameworks fail to capture the distinctive inter-cluster coupling inherent in multi-cluster HAPS-RSMA systems, where the common-stream power on each RB jointly impacts multiple clusters. To the best of our knowledge, GNN-based solutions tailored to the MMF-RSMA power allocation problem in multi-cluster HAPS networks remain largely unexplored. To bridge this gap, we propose a scalable heterogeneous GNN framework for MMF-RSMA power allocation (MMF--RSMA--PA) in HAPS networks.

\subsection{Contributions}\label{Contributions}
Motivated by these observations, we investigate interference management within the HAPS tier using UE clustering and RSMA. Unlike our previous works~\cite{Our-WCL,our-CL,our-ICC,TwoLevel}, where HAPS serves UEs through UE-specific beams, we consider a more practical transmission strategy in which the HAPS forms beams toward geographic areas, referred to as clusters, with multiple UEs served by the same beam. This assumption is motivated by the high altitude of HAPS, from whose perspective geographically separated UEs often exhibit only small angular separations.
In our earlier work~\cite{GCPaper}, we proposed an angular-aware UE clustering algorithm and formulated the corresponding MMF--RSMA--PA problem, which was solved using an iterative SCA algorithm. Although the SCA-based solution provides a high-quality benchmark, it must be solved from scratch for every network realization, resulting in prohibitive computational complexity for large-scale HAPS networks.
In this paper, we build upon that framework by representing the HAPS network as a heterogeneous graph~\cite{HetGraph} and developing a GNN-based power allocation policy. The proposed GNN directly predicts the RSMA power allocation, substantially reducing the computational complexity while achieving fairness performance comparable to the SCA benchmark and naturally generalizing to different network sizes.

The main contributions of this paper are summarized as follows.
\begin{itemize}
\item We propose a heterogeneous GNN policy for the MMF--RSMA--PA problem in multi-cluster HAPS networks. The HAPS network is represented as a heterogeneous graph comprising UE, cluster, and RB nodes, enabling the policy to capture the underlying communication topology.

\item We develop an unsupervised training framework that directly maximizes a smooth surrogate of the MMF objective. The proposed architecture incorporates a constraint-feasible decoder that maps the outputs to RSMA power allocations satisfying the total transmit-power and common-rate decodability constraints by construction.

\item Owing to the graph-based message-passing architecture, the proposed GNN is independent of the numbers of UEs, clusters, and RBs. Consequently, a single trained policy naturally generalizes to networks of different sizes without retraining. We further introduce an iterative rate-hint refinement mechanism that unrolls the allocation over a few forward passes, emulating the inner iterations of the SCA algorithm.

\item Extensive simulation results demonstrate that the proposed GNN achieves fairness performance comparable to the SCA-based benchmark in~\cite{GCPaper} while reducing the per-realization computation time by orders of magnitude and generalizing effectively across different network sizes.
\end{itemize}

The rest of the paper is organized as follows. Section~\ref{SystemModel} presents the system model, discusses the clustering algorithm, and formulates the MMF--RSMA--PA problem. Section~\ref{sec:graph} introduces the proposed heterogeneous graph representation of the HAPS multi-cluster network. Section~\ref{sec:gnn} presents the proposed GNN-based RSMA power allocation scheme. Numerical results are presented in Section~\ref{sec:results}, and Section~\ref{Sec:Conclusion} concludes the paper.

\section{System Model and Problem Formulation}\label{SystemModel}
\begin{figure}[t]
    \centering
    % \captionsetup{justification=centering}
    \includegraphics[width=0.5\linewidth]{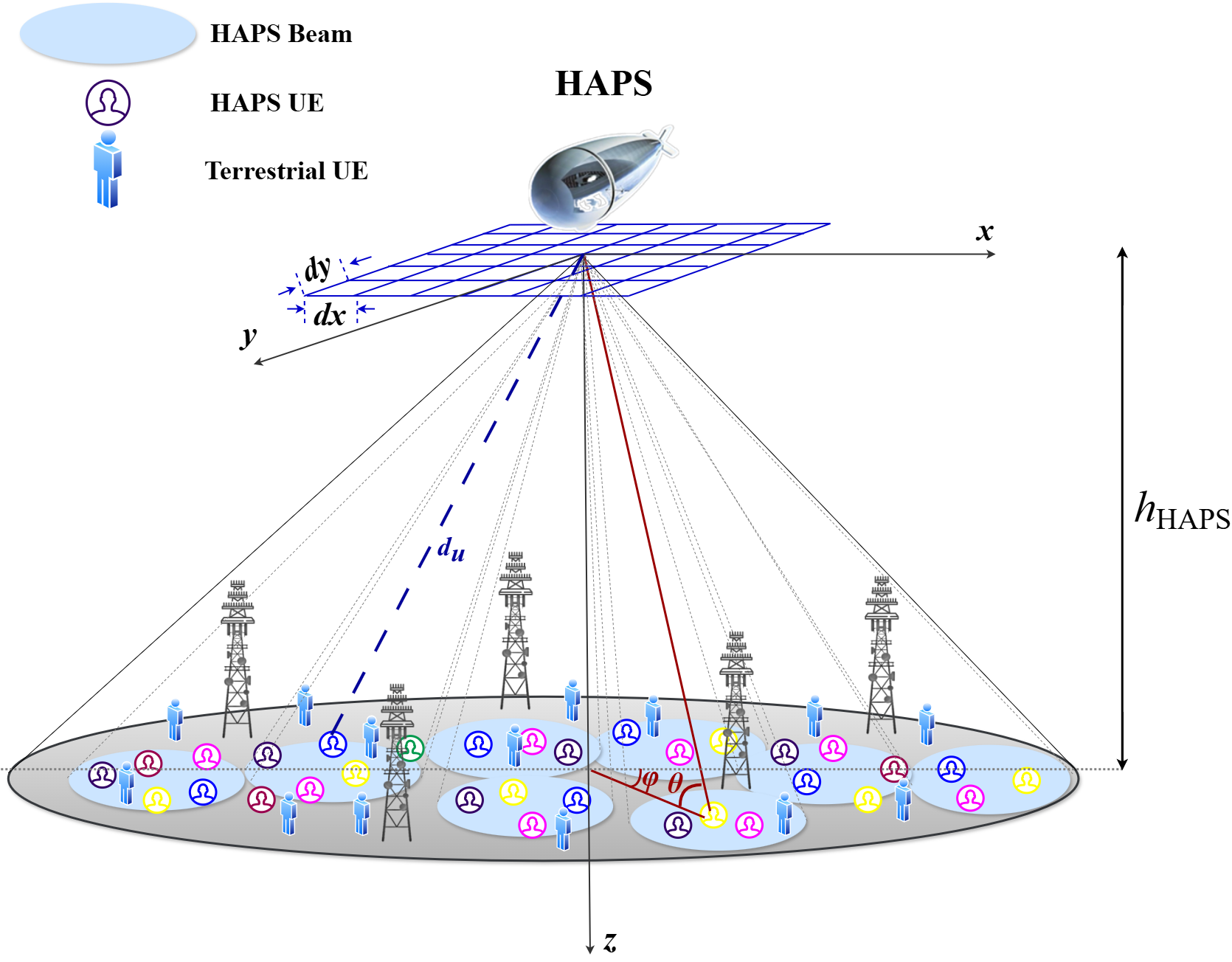}
    \caption{\small System model of the considered vHetNet, where the HAPS and terrestrial tiers operate over orthogonal frequency bands.}
    \label{systemmodel}
\end{figure}

We consider a vHetNet comprising a HAPS and multiple terrestrial macro base stations, operating in the downlink and covering a circular urban region with radius $A$ km. The HAPS complements the terrestrial network by using $R$ orthogonal RBs to serve $U$ single-antenna UEs experiencing insufficient terrestrial coverage or capacity, as illustrated in Fig.~\ref{systemmodel}. The HAPS operates over a dedicated \mbox{sub-6 GHz} spectrum that is orthogonal to the spectrum allocated to the terrestrial network. Consequently, inter-tier interference between the HAPS and terrestrial tiers is neglected. UEs and RBs are respectively indexed by the sets $\mathcal{U} \triangleq \{1, \dots, U\}$, with $u \in \mathcal{U}$, and $\mathcal{R} \triangleq \{1, \dots, R\}$, with $r \in \mathcal{R}$. The HAPS onboard antenna system consists of a uniform planar array (UPA) with $N_V \times N_H$ elements, where $N_V$ and $N_H$ represent the number of antenna elements along the vertical and horizontal directions, respectively.
Owing to the elevated altitude of the HAPS, propagation between the HAPS and each UE is predominantly governed by a LoS path. We accordingly model the large-scale channel deterministically, capturing only the LoS component. 

For a given UE $u$, we denote by $d_u$ the distance between the UE and HAPS, and by $(\theta_u,\phi_u)$ the corresponding elevation and azimuth departure angles from the HAPS, with $\theta_u \in [0,\pi/2]$ and $\phi_u \in [-\pi,\pi)$ (see Fig.~\ref{systemmodel}). 
Rather than forming a dedicated beam per UE, the HAPS generates directional beams pointed toward fixed geographic regions, referred to as clusters, each of which covers a unique group of UEs. Let $\mathcal{L}=\{1,\ldots,L\}$ denote the set of HAPS clusters, where each cluster, indexed by $\ell\in\mathcal{L}$, corresponds to a beam generated by the HAPS. The boresight direction of the beam corresponding to cluster $\ell$ is specified by the angle pair $(\theta_\ell, \phi_\ell)$. Accordingly, the channel gain experienced by UE $u$ under the beam of cluster $\ell$ is denoted $g_{\ell,u}$ and expressed as
\begin{align}
g_{\ell,u} = G_{\ell}(\theta_u,\phi_u) /\, \mathrm{PL}(d_u),~\forall \ell,~\forall u,\label{eq:channel_gain}
\end{align}
where $\mathrm{PL}(d_u)=\Big({4\pi d_u f_c}/{c}\Big)^2$ denotes the free-space path loss (FSPL) over distance $d_u$, with carrier frequency $f_c$ and speed of light $c$. The term $G_{\ell}(\theta_u,\phi_u)$ denotes the effective gain that cluster $\ell$'s beam provides in the direction of UE $u$'s angular location $(\theta_u,\phi_u)$; this quantity is derived from the combined radiation pattern of the HAPS UPA, which incorporates both the individual element pattern and the array factor as prescribed in the ITU recommendation~\cite[Tables 3 \& 4]{itur_m2101_2017}. The proposed interference management scheme consists of a clustering algorithm, followed by an RSMA-based transmission scheme, which are described in the following subsections.
\subsection{WU-Clustering Algorithm}
We define $\mathcal{U}_\ell$ to denote the set of UEs served by cluster $\ell$ where each UE $u$ belongs to exactly one cluster, i.e.,
$\mathcal{U}_\ell \cap \mathcal{U}_{\ell'} = \emptyset,\ \forall \ell \neq \ell'$.
Moreover, each UE $u$ is allocated a single RB, denoted by $r_u$. Since the number of RBs is limited ($R < U$), multiple UEs inevitably share the same RB, leading to intra-RB interference. To address this challenge, in our previous work \cite{GCPaper}, we proposed an angular-aware clustering and beam-steering framework that jointly performs UE clustering and beam design while allocating orthogonal RBs within each cluster. Owing to its effectiveness, the same worst-UE beam-steering clustering (WU-Clustering) algorithm is adopted in this work and is briefly reviewed below.
% efficient algorithm that jointly performs UE clustering and beam steering, and allocates orthogonal RBs within each cluster.Due to the effectiveness of the proposed angular-aware clustering with worst-UE beam steering (WU-Clustering), we will employ the same algorithm here. The algorithm has been explained briefly as follows.

\emph{\textbf{Review of WU-Clustering Algorithm \cite{GCPaper}:}} Since maximizing the minimum SE is the primary objective of both this work and our previous work~\cite{GCPaper}, the UE clustering algorithm is designed to support this objective. Due to the high altitude of HAPS, UEs with similar angular locations experience nearly identical beam gains. Consequently, co-scheduling such UEs on the same RB results in severe intra-RB inter-UE interference. To limit this effect, the proposed WU-Clustering Algorithm~1 in~\cite{GCPaper} relies purely on angular information and consists of two stages. First, it groups UEs into $L \triangleq \lceil U/R \rceil$ clusters based on their angular proximity. Then, it steers the beam of each cluster toward the UE with the worst channel condition to maximize the minimum beam gain within the cluster. 
Specifically, for every UE $u \in \mathcal{U}$, the angular feature vector is defined as
\begin{equation}\label{eq:angular_feature}
\mathbf{x}_u \triangleq [\theta_u,\ \cos\phi_u,\ \sin\phi_u]^{\top} \in \mathbb{R}^3,~\forall u,
\end{equation}
where representing the azimuth through $(\cos\phi_u,\sin\phi_u)$ rather than $\phi_u$ directly avoids the discontinuity at $\pm\pi$. Given $\mathbf{x}_u,~\forall u$, a capacity-constrained $K$-means procedure groups the $U$ UEs into $L$ clusters $\{\mathcal{U}_\ell\}_{\ell=1}^{L}$, where each cluster is limited to at most $R$ UEs so that its members can subsequently be assigned orthogonal RBs and intra-cluster interference can be avoided~\cite{Malinen2014}. For each cluster, the beam direction $(\theta_\ell,\phi_\ell)$ is then selected to maximize the gain of the worst-served UE as
\begin{equation}\label{BeamDirection}
    (\theta_\ell,\phi_\ell) \leftarrow \operatorname*{argmax}_{(\theta,\phi)} \min_{u \in \mathcal{U}_\ell} g_{\ell,u},~\forall \ell.
\end{equation}

After clustering,~\cite{GCPaper} proposed an interference-aware RB allocation approach, where RBs are assigned to minimize the aggregated leakage toward each newly scheduled UE. However, as shown in Table~\ref{tab:RB}, the performance improvement of this approach compared with random RB allocation is negligible. Therefore, in this work, we adopt random RB allocation within each cluster to simplify the overall framework without sacrificing performance. We denote the set of UEs assigned to RB $r$ by $\mathcal{U}_r$.
\begin{table}[h!]
    \centering
    \caption{\small Impact of RB allocation strategy on average SE.}
    \label{tab:RB}
    \small
    \begin{tabular}{|c|c|}
        \hline
        RB allocation scheme & Average SE (b/s/Hz/UE) \\
        \hline
        Interference-aware RB allocation~\cite{GCPaper} & 0.545 \\
        \hline
        Random RB allocation      & 0.538 \\
        \hline
    \end{tabular}
\end{table}

\subsection{MMF--RSMA--PA Problem Formulation}\label{Sec:RSMA_III}
The objective of the RSMA power allocation is to mitigate the inter-cluster interference experienced by UEs belonging to different clusters that share the same RB. To achieve this, the message intended for UE $u$, denoted by $W_u$, is divided into a common component $W_u^{(\mathrm{c})}$ and a private component $W_u^{(\mathrm{p})}$. For every RB $r$, the HAPS aggregates the common components of all UEs assigned to that RB, i.e., ${W_u^{(\mathrm{c})}: u \in \mathcal{U}_r}$, into a single RB-level common message $W_r^{(\mathrm{c})}$, which is encoded into the common stream $s_r^{(\mathrm{c})}$ with $\mathbb{E}[|s_r^{(\mathrm{c})}|^2]=1,~\forall r$. Meanwhile, the private message $W_u^{(\mathrm{p})}$ of each UE is independently encoded into its corresponding private stream $s_u^{(\mathrm{p})}$ with $\mathbb{E}[|s_u^{(\mathrm{p})}|^2]=1,~\forall u$.

Accordingly, the common stream $s^{(\mathrm{c})}_{r}$ of RB $r$ is transmitted simultaneously
over all $L$ cluster beams, where $p^{(\mathrm{c})}_{r,\ell},~\forall r\in\mathcal{R},~\forall\ell\in\mathcal{L}$ denotes the power
allocated to beam $\ell$ on that RB. In contrast, the private stream
$s^{(\mathrm{p})}_{u}$ is transmitted with power $p^{(\mathrm{p})}_{u}$ over the
beam of its serving cluster $\ell(u)$, on its assigned RB $r_{u}$, where
$\ell(u)$ denotes the index of the cluster serving UE $u$. Consequently, the
received signal at UE $u$ is expressed as
\begin{equation}\label{eq:rsma_rx}
\small
y_{u} = \underbrace{\sum_{\ell\in \mathcal{L}}\sqrt{p_{r_u,\ell}^{(\mathrm{c})} \, g_{\ell,u}} \, s_{r_u}^{(\mathrm{c})}}_{\text{received common part}}
+ \underbrace{\sum_{\substack{k\in\mathcal{U}_{r_u}}}\sqrt{p_{k}^{(\mathrm{p})} g_{\ell(k),u}}\, s_{k}^{(\mathrm{p})}}_{\text{received private part}}
+ n_{u},~\forall u, \qquad \qquad
\end{equation}
where $n_u\sim\mathcal{CN}(0,\sigma_n^2)$ represents the additive white Gaussian noise (AWGN) at UE $u$ with variance $\sigma_n^2$.

Each UE first decodes the common stream corresponding to its assigned RB while treating all private streams transmitted over the same RB as interference. Accordingly, the SINR for decoding the common stream at UE $u$ is given by
\begin{equation}\label{eq:sinr_common}
\gamma^{(\mathrm{c})}_{u}=
\frac{\sum\limits_{\ell\in \mathcal{L}} p_{r_u,\ell}^{(\mathrm{c})}\,  g_{\ell,u}}
{\sum\limits_{\substack{k\in\mathcal{U}_{r_u}}} p_{k}^{(\mathrm{p})} \, g_{\ell(k),u} + \sigma_\mathrm{n}^2},~\forall u,
\end{equation}
which leads to the achievable common rate as
\begin{equation}\label{eq:rate_common_user}
R^{(\mathrm{c})}_{u}=B \, \log_2\big(1+\gamma^{(\mathrm{c})}_{u}\big),~\forall u,
\end{equation}
where $B$ denotes the allocated bandwidth to each RB.
Since every UE sharing RB $r$ must successfully decode the common stream, the corresponding common transmission rate is constrained by the worst decoding UE, namely,
\begin{equation}\label{eq:rate_common}
R^{(\mathrm{c})}_{r}=\min\limits_{u\in\mathcal{U}_r} R^{(\mathrm{c})}_{u},~\forall r.
\end{equation}

Let $C_u\ge0$ denote the fraction of the common rate assigned to UE $u$. Then, the total allocated common rate on each RB must satisfy
\begin{equation}\label{eq:common_alloc}
\sum\limits_{u\in\mathcal{U}_r} C_u \le R_r^{(\mathrm{c})},~\forall r.
\end{equation}

After successfully decoding the common stream, UE $u$ removes it through SIC and subsequently decodes its own private stream while considering the remaining private streams on the same RB as interference. The resulting private-stream SINR is therefore
\begin{equation}\label{eq:sinr_private}
\gamma^{(\mathrm{p})}_{u}=
\frac{p^{(\mathrm{p})}_{u} g_{\ell(u),u}}
{\sum\limits_{\substack{k\in\mathcal{U}_{r_u}\\ k\neq u}} p_{k}^{(\mathrm{p})}\, g_{\ell(k),u}+\sigma_n^2},~\forall u,
\end{equation}
and the corresponding achievable private rate is
\begin{equation}\label{eq:rate_private}
R^{(\mathrm{p})}_{u}=B \, \log_2\big(1+\gamma^{(\mathrm{p})}_{u}\big),~\forall u.
\end{equation}

The adopted RSMA framework introduces additional flexibility for interference management by allocating common-stream power independently across cluster beams on every RB through the variables $p_{r,\ell}^{(\mathrm{c})},~\forall r\in\mathcal{R},~\forall\ell\in\mathcal{L}$. Nevertheless, the overall performance depends on an appropriate distribution of transmit power between common and private streams together with a suitable allocation of the common rate among UEs. Therefore, the variables $p_{r,\ell}^{(\mathrm{c})},~\forall r\in\mathcal{R},~\forall\ell\in\mathcal{L}$, $p_u^{(\mathrm{p})},~\forall u\in\mathcal{U}$, and $C_u,~\forall u\in\mathcal{U}$ are jointly optimized while satisfying the HAPS transmit power budget
\begin{equation}\label{eq:power_rsma}
\sum\limits_{r\in \mathcal{R}}\sum\limits_{\ell\in\mathcal{L}} p_{r,\ell}^{(\mathrm{c})}
+ \sum\limits_{u\in\mathcal{U}} p_{u}^{(\mathrm{p})}
  \le P_{\mathrm{T}},
  \end{equation}
  where $P_{\mathrm{T}}$ denotes the total transmit power available at the HAPS. The design objective is to jointly determine the common-stream powers, private-stream powers, and common-rate allocation so as to maximize the minimum achievable data rate among all UEs.

The overall achievable rate of UE $u$ is obtained by combining its allocated common rate and private rate, i.e.,
\begin{equation}\label{eq:user_total_rate}
R_{u}=C_u+R_u^{(\mathrm{p})},\quad \forall u.
\end{equation}

Accordingly, the resulting MMF--RSMA--PA problem is formulated as
\begin{subequations}\label{eq:mmf_problem}
\begin{align}
\small
\max_{\mathbf{P}^{(\mathrm{c})},\mathbf{p}^{(\mathrm{p})},\mathbf{C}} &~~ \min_{u\in\mathcal{U}} R_u \label{Objective_MMF}\\
\text{s.t.}\quad
& \eqref{eq:common_alloc},~\eqref{eq:power_rsma}, \\
& \mathbf{C}\ge 0, \mathbf{p}^{(\mathrm{p})}\ge 0, \mathbf{P}^{(\mathrm{c})}\ge 0, \label{Const:positive_Var}
\end{align}
\end{subequations}
where $\mathbf{P}^{(\mathrm{c})}$, $\mathbf{p}^{(\mathrm{p})}$, and $\mathbf{C}$ collect the optimization variables $p_{r,\ell}^{(\mathrm{c})},~\forall r\in\mathcal{R},~\forall\ell\in\mathcal{L}$, $p_u^{(\mathrm{p})},~\forall u\in\mathcal{U}$, and $C_u,~\forall u\in\mathcal{U}$, respectively. Constraint~\eqref{eq:common_alloc} guarantees that the common stream can be decoded by every UE assigned to the same RB, whereas constraint~\eqref{eq:power_rsma} limits the total transmit power consumed by the common and private streams to the available HAPS power budget. The optimization problem in \eqref{eq:mmf_problem} is non-convex because both the objective function in \eqref{Objective_MMF} and constraint~\eqref{eq:common_alloc} depend on the SINR expressions in \eqref{eq:sinr_common} and \eqref{eq:sinr_private}, which contain fractional functions of the optimization variables.
\begin{figure}
    \centering
    \includegraphics[width=0.7\linewidth]{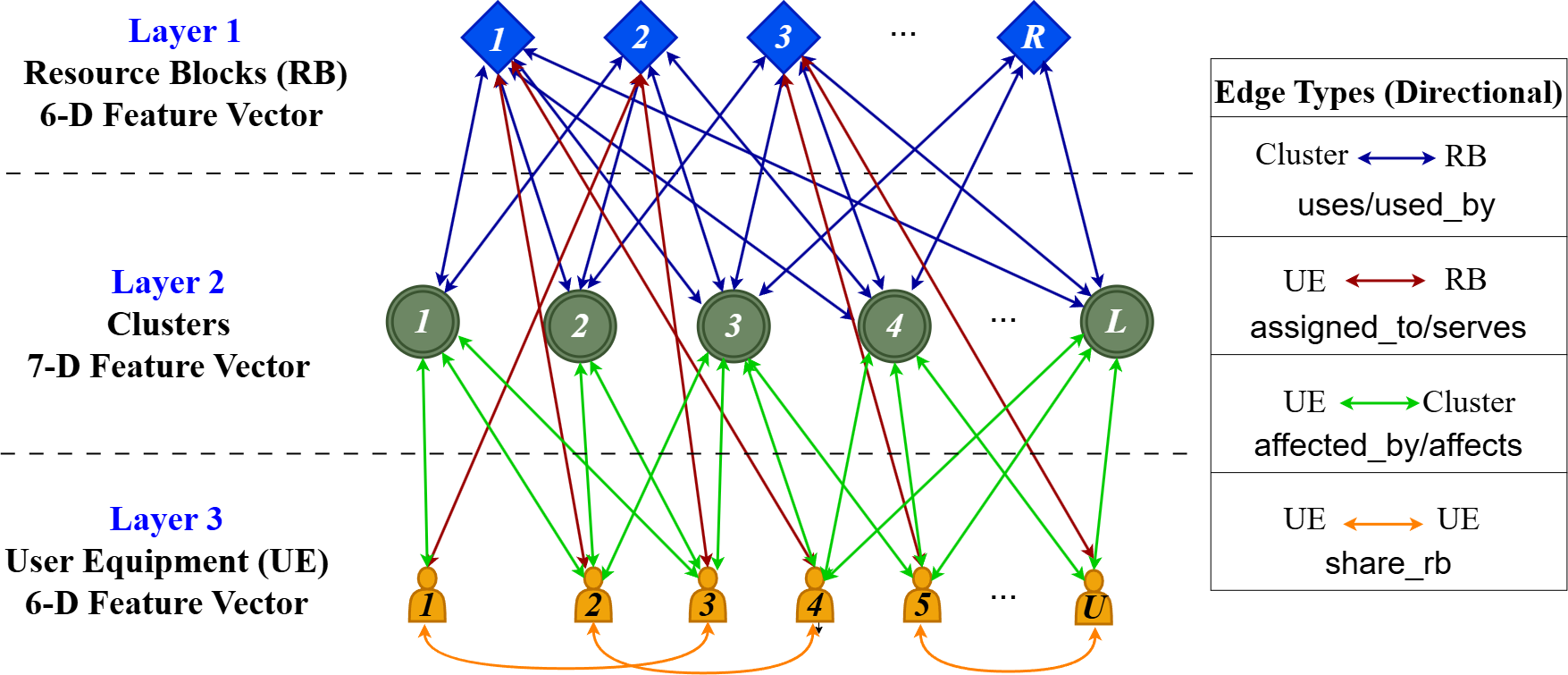}
    \caption{\small Heterogeneous graph representation of the HAPS network.}
    \label{Fig:Graph}
\end{figure}

While SCA can efficiently obtain a suboptimal solution to problem~\eqref{eq:mmf_problem} by iteratively solving a sequence of convex approximations, it must solve the optimization problem from scratch for every channel realization, without reusing information learned from previous UE topologies or cluster configurations. Furthermore, as the numbers of UEs and clusters increase, the problem dimensionality grows substantially, leading to higher computational complexity and longer convergence times. Consequently, SCA-based resource allocation is not suitable for a large-scale and real-time HAPS network, where resource allocation must adapt rapidly to dynamic UE distributions. To overcome these limitations, the following section proposes a GNN-based policy that directly learns the mapping from network realizations to RSMA power allocation decisions. Once trained, the proposed GNN outputs a feasible power allocation through a low-complexity forward pass, retaining the fairness performance of the SCA solution while operating at significantly lower computational cost.

\section{Graph Representation of the HAPS Network}
\label{sec:graph}
In this section, we represent the multi-cluster HAPS network as a heterogeneous graph~\cite{HetGraph}. Unlike homogeneous graphs, where all nodes and edges are assumed to be of the same type, heterogeneous graphs can naturally capture diverse node types and their interactions. Specifically, in the considered HAPS network, UEs, clusters, and RBs are modeled as distinct node types, while the corresponding edge types represent the different relationships among these nodes. This representation explicitly captures UE to cluster associations, RB allocations, and interference interactions, enabling the learning model to effectively exploit the underlying relational structure while remaining invariant to the arbitrary ordering of UEs, clusters, and RBs. 
Accordingly, each network realization is represented as a heterogeneous graph
$\mathcal{G}=(\mathcal{V},\mathcal{E})$, as shown in Fig.~\ref{Fig:Graph},
where the node set is defined as $\mathcal{V}=\mathcal{U} \cup \mathcal{L} \cup \mathcal{R}$,
consisting of nodes corresponding to three node types $\{\mathrm{ue},\mathrm{cl},\mathrm{rb}\}$ representing UEs, clusters, and RB nodes, respectively. The edge set is defined as
$\mathcal{E} = \{\text{UE}\leftrightarrow\text{UE},\,
\text{UE}\leftrightarrow\text{Cluster},\,
\text{UE}\leftrightarrow\text{RB},\,
\text{Cluster}\leftrightarrow\text{RB}\},$ consisting of edges corresponding to four edge-types, where each edge-type captures a specific physical interaction or association between network nodes. Furthermore, each node maintains a feature vector that captures its intrinsic characteristics, while each edge maintains relation-specific attributes that provide additional physical-layer information about the corresponding interaction. The details of the node features and relation-specific edge attributes are described in the following subsections.

\subsection{Node Features}
\label{sec:graph_nodes}
Each node type is described by a feature vector designed to capture the information most relevant to the MMF power-allocation problem. To improve training stability and prevent features with large numerical ranges from dominating the learning process, all node descriptors are standardized to zero mean and unit variance per realization. We denote the standardization operation by $\mathrm{z}(\cdot)$. While standardization removes scale variations and allows the model to focus on relative structure, it discards the absolute gain information, which is vital for determining the required transmit power levels. To retain this information, we append a single unstandardized log gain to every node feature. Moreover, all beam gains are mapped to the logarithmic domain through $(\cdot)^{\log}\!=\log_{10}(\cdot+\varepsilon)$, with a small floor $\varepsilon$ to avoid singularities.

\subsubsection{UE nodes} The common and private transmit powers allocated to UE $u$ depend on its location, the beam of its serving cluster $\ell(u)$, and the gains from the remaining clusters. Accordingly, each UE $u$ maintains the six-dimensional feature vector $\mathbf{f}^{\mathrm{ue}}_{u}\in\mathbb{R}^{6}$ formed by standardizing five descriptors and augmenting them with the unstandardized serving gain as
\begin{IEEEeqnarray*}{lcl}
\mathbf{f}^{\mathrm{ue}}_{u}=
\begin{bmatrix}
\mathrm{z}\!\left(\tilde{\mathbf{f}}^{\mathrm{ue}}_{u}\right)\\
g^\mathrm{log}_{\ell(u),u}
\end{bmatrix},~\forall u,\IEEEyesnumber \IEEEyessubnumber*\label{UE_Feature}\\
\tilde{\mathbf{f}}^{\mathrm{ue}}_{u}
=\Big[x_u,\, y_u,\, g^\mathrm{log}_{\ell(u),u},\, \hat{g}^\mathrm{log}_u\!, \,\bar{g}^\mathrm{log}_u\!\Big]^{\!\top},~\forall u,\label{eq:ue_feat_raw}
\end{IEEEeqnarray*}
where the first two descriptors $x_u$ and $y_u$ represent the normalized position of the UE $u$ in the coverage area (each normalized by the cell radius), $g_{\ell(u),u}$ denotes the gain of $\ell(u)$ towards UE $u$, $\hat{g}_u= \max_{\ell\in\mathcal{L}} g_{\ell,u}$ is the maximum cluster gain towards UE $u$, and $\bar{g}_u= \frac{1}{L}\sum_{\ell\in\mathcal{L}} g_{\ell,u}$ represents the mean gain over all clusters. These values together quantify how strongly the beams couple to the UE, i.e., its exposure to inter-cluster interference.

\subsubsection{Cluster nodes}
Each cluster $\ell$ maintains a seven-dimensional feature vector $\mathbf{f}^{\mathrm{cl}}_{\ell}\in\mathbb{R}^{7}$, consisting of six standardized descriptors of its served set $\mathcal{U}_{\ell}$, its centroid position and serving gain statistics, and its leakage characteristics, followed by the unstandardized mean serving gain. Accordingly, the feature vector of cluster $\ell$ can be expressed as
\begin{IEEEeqnarray*}{lcl}
\mathbf{f}^{\mathrm{cl}}_{\ell}=
\begin{bmatrix}
\mathrm{z}(\tilde{\mathbf{f}}^{\mathrm{cl}}_{\ell})\\
(\bar{g}^{\,\mathrm{in}}_\ell)^{\log}
\end{bmatrix},~\forall \ell,\IEEEyesnumber \IEEEyessubnumber*\label{eq:cl_feat}\\
\tilde{\mathbf{f}}^{\mathrm{cl}}_{\ell}=\Big[x_\ell,y_\ell,(\bar g^{\mathrm{in}}_\ell)^{\log},(\underline g^{\mathrm{in}}_\ell)^{\log},(\hat g^{\mathrm{in}}_\ell)^{\log},(\bar g^{\mathrm{out}}_\ell)^{\log}\Big]^{\!\top}\!,~\forall \ell,\qquad\label{eq:cl_feat_raw}
\end{IEEEeqnarray*}
where $(x_\ell,y_\ell)$ is the normalized position of the cluster centroid calculated as the mean position of the UEs served by cluster $\ell$. Furthermore, $\bar g^{\,\mathrm{in}}_\ell$, $\underline g^{\,\mathrm{in}}_\ell$, and $\hat g^{\,\mathrm{in}}_\ell$ denote the mean, minimum, and maximum serving gains over $u\in\mathcal U_\ell$, respectively, while $\bar{g}^{\,\mathrm{out}}_\ell$ denotes the mean gain of beam $\ell$ toward the UEs it does not serve, quantifying the interference it generates, defined as
\begin{equation}\label{eq:cl_leakage}
    \bar{g}^{\,\mathrm{out}}_\ell=\frac{1}{U-|\mathcal U_\ell|}\sum_{u\notin\mathcal U_\ell} g_{\ell,u},~\forall \ell.
\end{equation}

\subsubsection{RB nodes}
Each RB $r$ maintains a six-dimensional feature vector $\mathbf{f}^{\mathrm{rb}}_{r}\in\mathbb{R}^{6}$, consisting of five standardized descriptors of the UEs sharing RB $r$, followed by the unstandardized minimum serving gain,
\begin{IEEEeqnarray*}{lcl}\label{RB_Feature}
\mathbf{f}^{\mathrm{rb}}_{r}=
\begin{bmatrix}
\mathrm{z}(\tilde{\mathbf{f}}^{\mathrm{rb}}_{r})\\
(\underline{g}^{\mathrm{rb}}_{r})^{\log}
\end{bmatrix},~\forall r,\IEEEyesnumber \IEEEyessubnumber*\\
\tilde{\mathbf{f}}^{\mathrm{rb}}_{r}=\Big[
\tfrac{|\mathcal{U}_{r}|}{L},
(\underline{g}^{\mathrm{rb}}_{r})^{\log},
(\mu^{\mathrm{rb}}_{r})^{\log},(\hat{g}^{\mathrm{rb}}_{r})^{\mathrm{log}},(\chi_{r})^{\log}
\Big]^{\!\top}\!,~\forall r, \qquad 
\label{RB_Feature_unstandardized}
\end{IEEEeqnarray*}
where $\underline{g}^{\,\mathrm{rb}}_{r}$ and $\hat{g}^{\,\mathrm{rb}}_{r}$ denote the minimum and maximum serving gain among the UEs sharing RB $r$, respectively, $\mu^{\mathrm{rb}}_{r}$ denotes the mean serving gain on RB $r$, and $\chi_{r}$ represents the mean cross-cluster gain on RB $r$, serving as a measure of its interference potential.
\begin{IEEEeqnarray*}{lcl}\label{muandshao}
\mu^{\mathrm{rb}}_{r} = \frac{1}{|\mathcal{U}_{r}|} \sum_{u\in\mathcal{U}_{r}}g_{\ell(u),u},~\forall r, \IEEEyesnumber \IEEEyessubnumber* \label{mu}\\
\chi_{r} = \frac{1}{|\mathcal{U}_{r}|(L-1)} \sum_{u\in\mathcal{U}_{r}} \sum_{\ell\neq\ell(u)}g_{\ell,u},~\forall r.\label{shao}
\end{IEEEeqnarray*}

The minimum serving gain $\big(\underline{g}^{\,\mathrm{rb}}_{r}\big)^{\log}$ is retained both in standardized and unstandardized forms because it determines the per-RB common rate $R^{(\mathrm{c})}_r$ in \eqref{eq:rate_common}, and therefore provides an informative descriptor for the common-power allocation.

\subsection{Relations and Edge Attributes}
\label{sec:graph_edges}
The heterogeneous graph contains four edge types that model the relationships among UEs, clusters, and RBs. As described in the next section, the proposed GNN performs message passing along these edges. To enable bidirectional information exchange, each relation is represented by a pair of directed edges that share the same attributes.
\subsubsection{UE\,$\leftrightarrow$\,UE (\textsf{share\_rb})}
An edge exists between two UEs $u$ and $k$ if they are assigned to the same RB. Because RB allocation is orthogonal within each cluster, any such pair necessarily belongs to different clusters. Consequently, these edges capture the inter-cluster interference targeted by the proposed RSMA scheme. For the directed edge from UE $u$ to UE $k$, the edge attribute is defined as
\begin{equation}\label{eq:edgeattr_uu}
e^{\mathrm{uu}}_{u\rightarrow k}
=
\log_{10}\!\left(
\frac{g_{\ell(u),k}}{g_{\ell(k),k}}
\right),
\end{equation}
where $g_{\ell(u),k}$ is the gain from the serving beam of UE $u$ to UE $k$, while $g_{\ell(k),k}$ is the serving gain of UE $k$. This attribute quantifies the interference created by the serving beam of UE $u$ relative to the desired signal received by UE $k$.

\subsubsection{UE\,$\leftrightarrow$\,Cluster (\textsf{affected\_by}/\textsf{affects})}
Each UE $u$ is connected to every cluster that is active on its assigned RB $r_u$. Specifically, an edge exists between UE $u$ and cluster $\ell$ whenever there exists a UE $k\in\mathcal{U}_\ell$ with $r_k=r_u$. Therefore, UE $u$ is connected not only to its serving cluster but also to every interfering cluster active on the same RB. The corresponding edge attribute is
\begin{equation}\label{eq:edgeattr_uc}
\mathbf{e}^{\mathrm{uc}}_{u,\ell}
=
\Big[
g^{\log}_{\ell,u},
\;
\mathbb{1}\{\ell=\ell(u)\}
\Big]^{\!\top},~\forall u,~\forall \ell,
\end{equation}
where the first element is the logarithmic beam gain of cluster $\ell$ towards UE $u$, and $\mathbb{1}\{\cdot\}$ denotes the indicator function, which equals one if cluster $\ell$ serves UE $u$, and zero otherwise.

\subsubsection{UE\,$\leftrightarrow$\,RB (\textsf{assigned\_to}/\textsf{serves})}

Each UE is connected to its assigned RB $r_u$. Consequently, the edge carries the constant attribute
\begin{equation}\label{eq:edgeattr_ur}
e^{\mathrm{ur}}_{u,r_u}=1,~\forall u.
\end{equation}

\subsubsection{Cluster\,$\leftrightarrow$\,RB (\textsf{uses}/\textsf{used\_by})}
An edge exists between each cluster $\ell$ and each RB $r$, resulting in a complete bipartite graph over $\mathcal{L}$ and $\mathcal{R}$. The corresponding edge attribute is defined as
\begin{equation}
\label{eq:edgeattr_lr}
\mathbf{e}^{\mathrm{lr}}_{\ell,r}
=
\Big[
(\bar{g}_{\ell,r})^{\log},
\;
(\underline{g}_{\ell,r})^{\log}
\Big]^{\!\top},~\forall \ell,~\forall r,
\end{equation}
\begin{equation}
\label{eq:mean_gain_lr}
\bar{g}_{\ell,r}
=
\frac{1}{|\mathcal{U}_r|}
\sum_{u\in\mathcal{U}_r}
g_{\ell,u},~\forall \ell,~\forall r,
\end{equation}
\begin{equation}
\label{eq:min_gain_lr}
\underline{g}_{\ell,r}
=
\min_{u\in\mathcal{U}_r}
g_{\ell,u}\,,~\forall \ell,~\forall r,
\end{equation}
where the quantities $\bar{g}_{\ell,r}$ and $\underline{g}_{\ell,r}$ denote the mean and minimum beam gains from cluster $\ell$ to the UEs assigned to RB $r$, respectively. The complete bipartite construction follows directly from the proposed WU-Clustering algorithm, which ensures that each cluster serves at least one UE on every RB. Consequently, every cluster utilizes every RB, and the common-stream transmit power $p^{(\mathrm{c})}_{r,\ell}$ must be optimized for every cluster--RB pair $(\ell,r)\in\mathcal{L}\times\mathcal{R}$. Furthermore, according to \eqref{eq:sinr_common}, the common stream data rate of any UE $u\in\mathcal{U}_r$ depends on the aggregate common-signal contribution from all clusters.
This dependence justifies representing every cluster--RB pair in the graph, allowing the GNN to learn the corresponding common-power allocation.

\begin{figure*}
    \centering
    \includegraphics[width=0.9\linewidth]{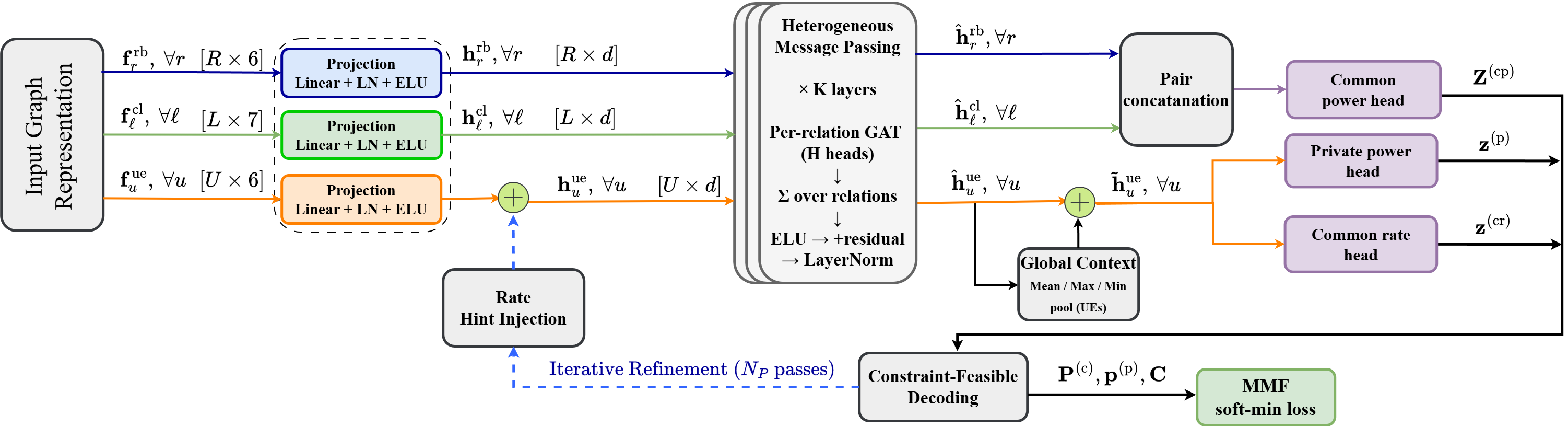}
    \caption{\small Proposed GNN architecture for MMF--RSMA--PA (GNN-RSMA).}
    \label{fig:GNNArch}
\end{figure*}

\section{Proposed GNN for MMF--RSMA--PA (GNN-RSMA)}
\label{sec:gnn}
In this section, we develop a GNN\footnote{To maintain brevity, an introduction to GNNs for wireless networks is omitted; interested readers are referred to \cite{Shen_GNN} for further details.} policy, parameterized by the weights $\boldsymbol{\theta}$, that maps a network realization directly to a feasible RSMA power allocation and common-rate allocation, $\big(\mathbf{p}^{(\mathrm{p})},\mathbf{P}^{(\mathrm{c})},\mathbf{C}\big)$, without any iterative optimization. The GNN is trained in an unsupervised manner by directly maximizing a smooth surrogate of the MMF objective function computed from its predicted allocation, thereby eliminating the need for labeled solutions generated by the SCA algorithm. The proposed architecture operates on the heterogeneous graph $\mathcal{G}$ defined in Section~\ref{sec:graph}. An overview of the proposed GNN architecture is illustrated in Fig.~\ref{fig:GNNArch}, and its individual components are described in the remainder of this section.

\subsection{Input Graph Representation and Node Embedding}
\label{sec:gnn_input}
 In the heterogeneous graph $\mathcal{G}$, different node types (UEs
\footnote{Although developed for a HAPS network with ground UEs, the proposed GNN-RSMA algorithm can be extended to networks involving flying UEs, such as uncrewed aerial vehicles (UAVs).}
, clusters, and RB nodes) have different feature dimensions and therefore cannot be processed directly by a common message-passing operator. To obtain a unified latent representation, the feature vector of each node $v$ is projected into a common $d$-dimensional embedding space using a projection layer tailored to its node type. This layer consists of a linear transformation, layer normalization (LN), and exponential linear unit (ELU) activation. Specifically, for each node $v\in\mathcal{V}$, the resulting $d$-dimensional embedding, denoted by $\mathbf{h}^{\tau(v)}_v$, is given by
\begin{equation}
\label{eq:node_encoder}
\mathbf{h}^{\tau(v)}_v
=
\mathrm{ELU}\big(\mathrm{LN}(\mathbf{W}_{\tau(v)}\mathbf{f}^{\tau(v)}_v + \mathbf{b}_{\tau(v)})\big),
\end{equation}
where $\tau(v) \in \{\mathrm{ue},\mathrm{cl},\mathrm{rb}\}$ denotes the node type of $v$, $\mathbf{f}^{\tau(v)}_v$ is the corresponding input feature vector, and $\mathbf{W}_{\tau(v)}$, $\mathbf{b}_{\tau(v)}$ are the weight matrix and bias associated with the projection layer corresponding to that node type. 
A single linear layer is intentionally used instead of a deeper multi-layer perceptron (MLP), since the purpose of this stage is only to project heterogeneous node features, whose dimensions vary across node types, into a common $d$-dimensional embedding space. The extraction of higher-level nonlinear representations and the modeling of interactions among nodes are subsequently performed by the message-passing layers.
LN stabilizes the scale of the projected embeddings across heterogeneous nodes, whose raw features can differ substantially in magnitude and distribution, preventing any single node type from dominating the subsequent message-passing computation due to scale imbalance. ELU then supplies the nonlinearity needed for expressive initial embeddings while preserving smooth, non-zero gradients for negative inputs, improving gradient flow relative to alternatives such as ReLU. Together, this projection maps all node types into a common latent space while preserving type-specific feature extraction, enabling a unified heterogeneous message-passing procedure.
\subsection{Heterogeneous Message Passing}\label{sec:messagepassing}
\label{sec:gnn_mp}
After the initial node embeddings have been obtained, message passing enables each node to aggregate information from its graph neighborhood so that the final embeddings capture the interactions among UEs, clusters, and RBs. The message-passing module consists of $K$ layers. Since different edge-types represent different physical interactions, each edge $\rho\in\{\mathrm{uu},\mathrm{uc},\mathrm{ur},\mathrm{lr}\}$ is processed by an independent edge-conditioned graph attention network (GAT) operator, allowing the model to learn a distinct attention mechanism for each type of interaction rather than treating all relations identically. For each relation $\rho$, multi-head attention with $H$ heads is employed, indexed by $h\in\{1,\ldots,H\}$, to allow the model to jointly attend to different aspects of the neighborhood.

For each node $v$ and a relation $\rho$, let $\mathcal{N}_{\rho}(v)$ denote the set of neighboring nodes connected to $v$ through relation $\rho$. The attention score for a neighbor $w\in\mathcal{N}_{\rho}(v)$ under head $h$ is computed as
\begin{equation}
s^{\rho,(h)}_{vw}
=
\sigma_{\!L}\!\left(
\mathbf{a}_{\rho}^{(h)\top}
\begin{bmatrix}
\mathbf{W}_{\rho}^{(h)}\mathbf{h}^{\tau(v)}_{v}\\
\mathbf{W}_{\rho}^{(h)}\mathbf{h}^{\tau(w)}_{w}\\
\mathbf{U}_{\rho}^{(h)}\mathbf{e}^{\rho}_{wv}
\end{bmatrix}
\right),
\label{eq:gat_score}
\end{equation}
where $\mathbf{e}^{\rho}_{wv}$ is the attribute of the directed edge $w\rightarrow v$, $\sigma_L(\cdot)$ is the LeakyReLU activation function, and $\mathbf{W}_{\rho}^{(h)}$, $\mathbf{U}_{\rho}^{(h)}$, and $\mathbf{a}_{\rho}^{(h)}$ are learnable parameters specific to relation $\rho$ and head $h$. Including the edge attribute $\mathbf{e}^{\rho}_{wv}$ in the attention score allows the model to weigh each neighbor not only by its state but also by the physical link condition it represents, such as channel gain or interference level. The normalized attention score is then obtained as
\begin{equation}
\alpha^{\rho,(h)}_{vw}
=
\operatorname*{softmax}_{w\in\mathcal{N}_{\rho}(v)}
(s^{\rho,(h)}_{vw}).
\label{eq:gat}
\end{equation}

Accordingly, the relation-specific message received by node $v$ under head $h$ is given by
\begin{equation}
\mathbf{m}^{\rho,(h)}_{v}
=
\sum_{w\in\mathcal{N}_{\rho}(v)}
\alpha^{\rho,(h)}_{vw}
\mathbf{W}_{\rho}^{(h)}\mathbf{h}^{\tau(w)}_{w},
\end{equation}
and the $H$ head outputs are averaged to obtain the relation-specific message
\begin{equation}
\mathbf{m}^{\rho}_{v}
=
\frac{1}{H}\sum_{h=1}^{H}
\mathbf{m}^{\rho,(h)}_{v},
\end{equation}
which stabilizes training by reducing the variance introduced by any single attention head. The messages from all relations connected to node $v$ are then combined, and the node embedding is updated using a residual connection, ELU activation, and layer normalization:
\begin{equation}
\small
\mathbf{h}^{{\tau(v)}}_{v,k}
=
\mathrm{LN}
\left(
\mathrm{ELU}
\left(
\sum_{\rho}
\mathbf{m}^{\rho}_{v}
\right)
+
\mathbf{h}^{{\tau(v)}}_{v,(k-1)}
\right),
\label{eq:update}
\end{equation}
where $k\in \{1,\ldots,K\}$ denotes the message-passing layer index. The residual connection allows each layer to refine the existing representation rather than replace it, which improves training stability and mitigates over-smoothing, a common issue in which repeated neighborhood aggregation makes node embeddings across the network indistinguishable. We denote the node embedding after $K$ message-passing layers by $\hat{\mathbf{h}}^{\tau(v)}_v$, i.e., $\hat{\mathbf{h}}^{\tau(v)}_v = \mathbf{h}^{{\tau(v)}}_{v,K}$. Note that with $K$ message-passing layers, each node embedding incorporates information from nodes up to $K$ hops away in the graph, which is sufficient to capture the multi-hop UE--cluster--RB interactions relevant to resource allocation. Because the message-passing functions are shared across nodes and operate only on local graph neighborhoods rather than fixed-size tensors, the trained GNN can be applied directly to network realizations with different numbers of UEs, clusters, and RBs without retraining or architectural changes.

\subsection{Global Context and Output Heads}
\label{sec:gnn_heads}
Message passing captures only local interactions within a limited number of hops, whereas the MMF objective is inherently global, as it is determined by the minimum SE across all UEs in the network. To provide each UE with global information, a context vector is extracted after the message-passing layers using mean, element-wise maximum, and element-wise minimum pooling of the final UE embeddings. This context vector is then injected into every UE embedding as
\begin{equation}\label{eq:context}
\small
\tilde{\mathbf{h}}^{\mathrm{ue}}_{u}
=
\hat{\mathbf{h}}^{\mathrm{ue}}_{u}
+
\mathrm{ELU}\!\left(
\mathrm{LN}\!\left(
\mathbf{W}_{\mathrm{ctx}}
\begin{bmatrix}
\frac{1}{U}\sum_{u\in\mathcal{U}}\hat{\mathbf{h}}^{\mathrm{ue}}_{u}\\
\max_{u\in\mathcal{U}}\hat{\mathbf{h}}^{\mathrm{ue}}_{u}\\
\min_{u\in\mathcal{U}}\hat{\mathbf{h}}^{\mathrm{ue}}_{u}
\end{bmatrix}
+
\mathbf{b}_{\mathrm{ctx}}
\right)
\right),
\end{equation}
where the maximum and minimum operations are performed element-wise over all UE embeddings, and $\mathbf{W}_{\mathrm{ctx}}$, $\mathbf{b}_{\mathrm{ctx}}$ are the learnable weight and bias of the context projection. The mean pooling captures the average network condition, while the maximum and minimum pooling provide information about extreme UE conditions. In particular, the minimum-pooled representation provides an explicit indication of the bottleneck UE, which helps the policy allocate resources according to the MMF objective rather than maximizing aggregate performance.

Three MLP heads are then used to generate the allocation variables from the context-aware embeddings. Each head is a two-layer feedforward network with an ELU nonlinearity, and outputs unconstrained scalar logits representing relative preferences rather than physical quantities directly. The private-power head and the common-rate head operate on each UE embedding $\tilde{\mathbf{h}}^{\mathrm{ue}}_u$ and produce the logits $z^{(\mathrm{p})}_u$ and $z^{\mathrm{(cr)}}_u$, respectively. The common-power head operates on each cluster--RB pair and generates
\begin{equation}
z^{(\mathrm{cp})}_{\ell,r}
= \mathrm{MLP}_{c}
\left(
\begin{bmatrix}
\hat{\mathbf{h}}^{\mathrm{cl}}_{\ell}\\
\hat{\mathbf{h}}^{\mathrm{rb}}_{r}
\end{bmatrix}
\right),~\forall
(\ell,r)\in\mathcal{L}\times\mathcal{R},
\end{equation}
where the cluster and RB embeddings used are the outputs of the message-passing stage, since the global context is injected only into the UE embeddings and is not required for setting the per-cluster, per-RB common power. The resulting logits $z^{(\mathrm{p})}_u$, $z^{\mathrm{(cr)}}_u$, and $z^{(\mathrm{cp})}_{\ell,r}$ are not directly interpreted as physical powers or rates; instead, they are transformed into a feasible RSMA allocation in the subsequent decoding step.

\subsection{Constraint-Feasible Decoding}
\label{sec:gnn_decode}
A central design choice is to enforce the constraints in \eqref{eq:mmf_problem} through a fixed differentiable decoder rather than penalty terms in the loss function. This guarantees feasibility throughout training and allows the GNN to learn only relative allocation preferences instead of learning the constraints implicitly. First, all power logits are concatenated and mapped through a single softmax scaled by the total power budget $P_T$ as
\begin{equation}
\big[\mathbf{p}^{(\mathrm{p})};\operatorname{vec}(\mathbf{P}^{(\mathrm{c})})\big]
=
P_T\,\operatorname{softmax}\!\big(
[\mathbf{z}^{(\mathrm{p})};
\operatorname{vec}(\mathbf{Z}^{(\mathrm{cp})})]
\big),
\label{eq:gnn_power}
\end{equation}
which produces non-negative powers that sum exactly to $P_T$. Therefore, the total-power constraint in \eqref{eq:power_rsma} is always satisfied with equality. Since private and common powers compete within the same softmax normalization, the policy automatically learns their tradeoff under the shared power budget.
Given the resulting powers, the common-stream SINR and the per-UE common rate are computed according to \eqref{eq:sinr_common}--\eqref{eq:rate_common}. Since the common message on a given RB must be decodable by every UE scheduled on that RB, through constraint \eqref{eq:common_alloc}, the resulting common rate of each RB is then distributed among its scheduled UEs through a per-RB softmax applied to the common-rate logits,
\begin{equation}
C_u
=
\Big(
\operatorname*{softmax}_{v\in\mathcal{U}_{r_u}}
z^{\mathrm{(cr)}}_v
\Big)_u
R^{(\mathrm{c})}_{r_u},~\forall u,
\label{eq:gnn_split}
\end{equation}
where $R^{(\mathrm{c})}_{r_u}$ is the common rate of the RB assigned to UE $u$. Because the softmax in \eqref{eq:gnn_split} is computed only over the UEs sharing the same RB, the resulting shares sum to one within each group, and hence $\sum_{u\in\mathcal{U}_r}C_u=R^{(\mathrm{c})}_r$, satisfying the decodability constraint in \eqref{eq:common_alloc} with equality. Finally, the private SINR and the total achievable rate are calculated according to \eqref{eq:sinr_private}--\eqref{eq:rate_private}.

\subsection{Iterative Refinement}
\label{sec:gnUnroll}
A key challenge in a single forward pass is that power allocation and bottleneck-UE identification are interdependent. Specifically, powers must be allocated before UE rates can be evaluated, while the bottleneck-UE can only be identified from those rates. Therefore, the initial policy output is generated without explicit knowledge of the bottleneck UE.
In the SCA algorithm, this coupling is resolved through iterative optimization. To overcome this limitation, we emulate an iterative refinement process by unrolling the GNN policy over $N_{\mathrm{p}}$ refinement passes, resulting in $N_{\mathrm{p}}+1$ forward passes in total. The same GNN, with shared parameters across all passes, is repeatedly applied to the graph, allowing the power-allocation decisions to be progressively refined as more information becomes available.
The first pass is executed without any rate hint, which is equivalent to leaving the UE embeddings from \eqref{eq:node_encoder} unchanged. After each pass, the achieved UE rates are used to form the rate hint for the next pass. Specifically, before message passing in every subsequent pass $n\in\{2,\ldots,N_{\mathrm{p}}+1\}$, the rate hint is injected into the UE embedding as
\begin{equation}
\mathbf{h}^{\mathrm{ue}}_{u}
\leftarrow
\mathbf{h}^{\mathrm{ue}}_{u}
+
\phi\!\big(\log(1+\hat{R}_u^{(n-1)})\big),~\forall u,
\label{eq:hint}
\end{equation}
where $\hat{R}_u^{(n-1)}$ is the rate achieved by UE $u$ in the previous pass, and $\phi(\cdot)$ is a learnable projection network consisting of a linear layer, layer normalization, and an ELU activation, which maps the scalar rate hint to the hidden feature dimension. The logarithmic mapping compresses the dynamic range of the rate hints while preserving their relative differences. Consequently, from the second pass onward, the embeddings carry information about the rates obtained in the previous pass, allowing the policy to refine the power allocation in subsequent passes. To reduce the training cost, the first $N_{\mathrm{p}}$ passes are executed without gradient tracking and are used only to generate progressively refined rate hints. The final pass is fully differentiable and is used to evaluate the training loss. As a result, the memory and computational cost of back-propagation remain independent of $N_{\mathrm{p}}$, while still benefiting from iterative refinement. Since the same GNN parameters are shared across all passes, optimizing the final pass simultaneously updates the policy used at every refinement step. Setting $N_{\mathrm{p}}=0$ reduces the method to a single forward pass without refinement.
\begin{algorithm}[t]
\caption{\small GNN training algorithm for MMF--RSMA--PA.}
\label{alg:gnn}
\begin{algorithmic}[1]
\small
\State \textbf{Input:} batch size $L_\mathrm{B}$, number of refinement passes $N_{\mathrm{p}}$, total power $P_T$, learning rate, soft-floor penalty $\lambda_{\mathrm{fl}}$, rate threshold $R_{\mathrm{th}}$, soft-min sharpness $\beta$.
\State \textbf{Output:} trained parameters $\boldsymbol{\theta}^{\star}$.
\State Initialize $\boldsymbol{\theta}$ and construct a validation set.
\Repeat
  \State Update $\beta$ according to its annealing schedule.
  \State Generate $L_\mathrm{B}$ independent network realizations.
  \State Compute channel gains $\mathbf{G}$ and construct graphs $\{\mathcal{G}_b\}_{b=1}^{L_\mathrm{B}}$.
  \State Initialize the rate hint $\hat{\mathbf{R}}\leftarrow\varnothing$ (no injection in the first pass).
  \For{$n=1,\ldots,N_{\mathrm{p}}$}
    \State (No grad.) Run policy network:
    \Statex \hspace{4.6em}$(\mathbf{z}^{(\mathrm{p})},\mathbf{z}^{(\mathrm{cr})},\mathbf{Z}^{(\mathrm{cp})})\leftarrow\pi_{\boldsymbol{\theta}}(\{\mathcal{G}_b\}_{b=1}^{L_\mathrm{B}},\hat{\mathbf{R}})$.
    \State Decode allocation and compute rates $\mathbf{R}$.
    \State Update hint $\hat{\mathbf{R}}\leftarrow\mathbf{R}$.
  \EndFor
  \State (Grad.) Final forward pass:
  \Statex \hspace{4.6em}$(\mathbf{z}^{(\mathrm{p})},\mathbf{z}^{(\mathrm{cr})},\mathbf{Z}^{(\mathrm{cp})})\leftarrow\pi_{\boldsymbol{\theta}}(\{\mathcal{G}_b\}_{b=1}^{L_\mathrm{B}},\hat{\mathbf{R}})$.
  \State Decode $(\mathbf{p}^{(\mathrm{p})},\mathbf{P}^{(\mathrm{c})},\mathbf{C})$ using \eqref{eq:gnn_power} and \eqref{eq:gnn_split}.
  \State Compute rates $\mathbf{R}$ and the batch loss $\mathcal{L}=\frac{1}{L_\mathrm{B}}\sum_{b=1}^{L_\mathrm{B}}\mathcal{L}_b$.
  \State Back-propagate to obtain $\nabla_{\boldsymbol{\theta}}\mathcal{L}$; clip the gradient.
  \State Update $\boldsymbol{\theta}$ using Adam; update the learning rate.
  \State Evaluate on the validation set and save the model.
\Until{convergence}
\end{algorithmic}
\end{algorithm}

\subsection{Unsupervised Training}
\label{sec:gnn_train}
The GNN policy is trained by directly maximizing the achievable MMF objective. However, the original objective $\min_{u\in\mathcal{U}}R_u$ is not well suited for gradient-based optimization because its gradient depends only on the single worst-performing UE, resulting in a sparse and unstable training signal. To obtain a smoother objective, we replace the minimum rate with the following log-sum-exp based lower bound: 
\begin{equation}
\widetilde{R}_{\min}(\boldsymbol{\theta})
=
-\frac{1}{\beta}
\log
\sum_{u\in\mathcal{U}}
e^{-\beta R_u},
\label{eq:softmin}
\end{equation}
where $\beta$ controls the approximation sharpness. The gradient of this surrogate is distributed among all UEs while assigning larger weights to UEs with lower rates. During training, $\beta$ is gradually increased so that $\widetilde{R}_{\min}$ approaches the true minimum rate as the UE rates become more balanced. The training loss further includes a soft-floor penalty that provides additional gradient when a UE rate falls below a threshold $R_{\mathrm{th}}$,
\begin{equation}\label{eq:gnn_loss}
\mathcal{L}(\boldsymbol{\theta})
=
-\widetilde{R}_{\min}(\boldsymbol{\theta})
+
\lambda_{\mathrm{fl}}
\sum_{u\in\mathcal{U}}
(R_{\mathrm{th}}-R_u)_{+},
\end{equation}
where $(\cdot)_+=\max(\cdot,0)$ and $\lambda_{\mathrm{fl}}$ controls the penalty weight. Instead of relying on a fixed dataset, each gradient update minimizes the average loss over a mini-batch of newly generated network realizations, with size $L_\mathrm{B}$, following standard mini-batch stochastic optimization. This estimates the expected loss $\mathbb{E}_{\mathcal{G}}[\mathcal{L}]$ and improves generalization across different UE distributions and channel realizations. The model parameters $\boldsymbol{\theta}$ are optimized using Adam with cosine learning-rate decay and gradient clipping for stable training. The model achieving the best validation MMF rate is retained. The complete training procedure is summarized in Algorithm~\ref{alg:gnn}.
% ======================================================================
%  Section: Numerical Results -- Simulation Setup (first subsection)
%
%  Requires: amsmath, amssymb.
%  Cross-references to remap to your labels:
%    \ref{tab:SimulationParameters}  (your parameters table)
%    \ref{sec:graph}, \ref{sec:gnn_train}, \ref{sec:gnUnroll}
%    \ref{sec:system}, \ref{sec:cluster}, \ref{sec:mmf}
%    \cite{cvx}, \cite{mosek}  (CVX / MOSEK, if you cite the solver)
% ======================================================================

\section{Simulation Results and Discussion}
\label{sec:results}
In this section, we report the results of simulations conducted to evaluate the effectiveness of the proposed GNN-based MMF--RSMA--PA (GNN-RSMA) algorithm. We begin by describing the simulation setup for the HAPS network and the proposed GNN architecture.

\textbf{\emph{HAPS Network Setup:}} We consider a single HAPS at an altitude of $20$~km, serving $U=60$ single-antenna UEs uniformly distributed over a circular area with radius $2$~km. The HAPS is equipped with an $8\times8$ UPA and serves the UEs over $R=10$ orthogonal RBs. The WU-Clustering scheme described in Section~\ref{SystemModel} is applied to determine the cluster assignments and beam directions. The effective antenna gains $g_{\ell,u}$ are obtained from the composite UPA radiation pattern based on the ITU model~\cite{itur_m2101_2017}. The remaining simulation parameters and their values are summarized in Table~\ref{tab:SimulationParameters}.
\begin{table}[!t]
\caption{Simulation parameters.}\label{tab:SimulationParameters}
\centering
\begin{tabular}{|c||c|}
\hline
\textbf{Parameter} & \textbf{Value}\\
\hline
\multicolumn{2}{|c|}{\textbf{System / channel}}\\
\hline
HAPS altitude & $20$ km\\
\hline
Center frequency ($f_c$) & $2.545$ GHz~\cite{itu_wrc}\\
\hline
$d_x,~d_y$ & $\lambda/2$\\
\hline
$\sigma^2_n$ & $-100$ dBm\\
\hline
$N_{\mathrm{V}} \times N_{\mathrm{H}}$ & $8 \times 8$\\
\hline
$P_{T}$ & $55$ dBm~\cite{itu_wrc}\\
\hline
HAPS antenna element $3$ dB beamwidth & $65$ degrees~\cite{itu_wrc}\\
\hline
HAPS antenna element gain & $8$ dBi~\cite{itu_wrc}\\
\hline
HAPS antenna element front to back ratio & $30$ dB~\cite{itu_wrc}\\
\hline
\multicolumn{2}{|c|}{\textbf{GNN architecture}}\\
\hline
Hidden dimension ($d$) & $256$\\
\hline
Message-passing layers ($K$) & $5$\\
\hline
GAT attention heads ($H$) & $4$\\
\hline
Node feature dim. (UE / cluster / RB) & $6$ / $7$ / $6$\\
\hline
Refinement passes ($N_{\mathrm{p}}$) & $4$\\
\hline
\multicolumn{2}{|c|}{\textbf{GNN training}}\\
\hline
Optimizer & Adam\\
\hline
Learning rate (initial) & $3\times10^{-4}$\\
\hline
Weight decay & $10^{-5}$\\
\hline
LR schedule & cosine annealing\\
\hline
Minimum learning rate & $10^{-6}$\\
\hline
Mini-batch size ($L_\mathrm{B}$) & $32$\\
\hline
Gradient-norm clip & $10$\\
\hline
Soft-min sharpness ($\beta$) & $10 \to 30$ (annealed)\\
\hline
SE threshold ($R_\mathrm{th}/B$), $\lambda_{\mathrm{fl}}$ & $0.4$,~$1$\\
\hline
Validation set size & $200$\\
\hline
\end{tabular}
\end{table}

\textbf{\emph{GNN Architecture Setup:}} Each network realization is represented by the heterogeneous graph introduced in Section~\ref{sec:graph}, where the UE, cluster, and RB node features have dimensions $6$, $7$, and $6$, respectively. The GNN policy employs a hidden dimension of $d=256$, $K=5$ heterogeneous message-passing layers, and $H=4$ attention heads for each relation. The number of refinement passes is set to $N_{\mathrm{p}}=4$ as described in Section~\ref{sec:gnUnroll}. The model is trained for $100,000$ epochs, in an unsupervised manner using Algorithm~\ref{alg:gnn}. In each training epoch, the loss is averaged over a mini-batch of $L_\mathrm{B}=32$ newly generated network realizations. Therefore, the policy learns the underlying resource allocation mapping rather than memorizing a fixed dataset. We use the Adam optimizer with parameters summarized in Table~\ref{tab:SimulationParameters}. The soft-min parameter $\beta$ is gradually increased from $10$ to $30$ during training to progressively tighten the surrogate toward the true minimum rate. The proposed GNN is trained and run on a system with an Intel Core Ultra 9 CPU and an NVIDIA GeForce RTX 5070 Laptop GPU in PyTorch using the PyTorch Geometric library.

The performance of the proposed scheme is compared against the SCA-based MMF--RSMA--PA algorithm (SCA-RSMA), proposed in~\cite{GCPaper}, which serves as an optimization benchmark. For each realization, the SCA problem is solved using the CVX toolbox with \texttt{mosek} 9.1.9 as an internal solver~\cite{CVX}. Both approaches are evaluated over the same set of $1000$ independent random realizations. 
Unless otherwise stated, for each network realization, the minimum SE among all UEs (i.e., $\min_u R_u/B$), denoted by $\mathrm{SE}_{\min}$, is considered as the performance metric. This metric represents the MMF objective function and characterizes the performance of the worst-UE.
We also report Jain's fairness index, defined as
\begin{equation}
J(\mathbf{R})=\frac{\big(\sum_{u\in\mathcal{U}}R_u\big)^{2}}{U\sum_{u\in\mathcal{U}}R_u^{2}},
\label{eq:jain}
\end{equation}
where $J(\mathbf{R})\in[1/U,1]$ and $J(\mathbf{R})=1$ indicates identical rates among all UEs. Moreover, the computational complexity is evaluated by comparing the per-realization inference time of the proposed GNN-RSMA with $N_{\mathrm{p}}$ refinement passes against the iterative SCA solver. 
\begin{figure}
    \centering
    \includegraphics[width=0.7\linewidth]{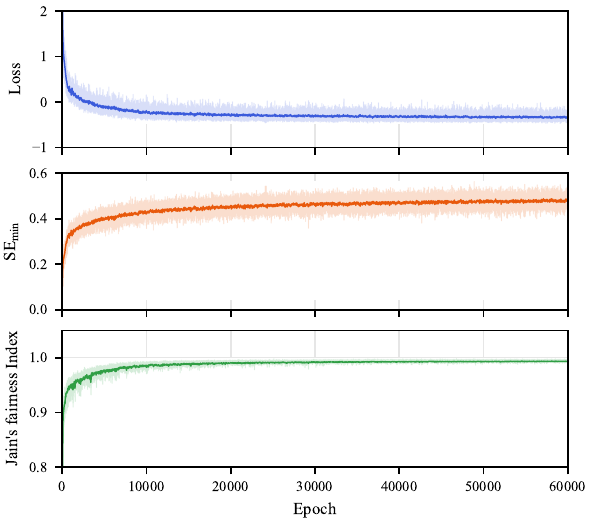}
    \caption{\small Convergence behavior of the proposed GNN-RSMA, trained using \mbox{Algorithm~\ref{alg:gnn}}.}
    \label{fig:Convergence}
\end{figure}

\begin{figure}[t]
    \centering
    \includegraphics[width=0.7\linewidth]{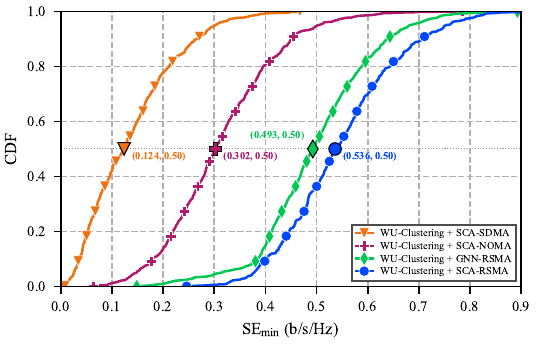}
    \caption{\small CDF of $\textrm{SE}_\textrm{min}$ for the proposed GNN-RSMA and the SCA-based RSMA, NOMA, and SDMA baselines.}
    \label{fig:cdf}
\end{figure}
\subsection{Convergence Behavior}\label{SubSec:Convergence}
Fig.~\ref{fig:Convergence} illustrates the training convergence of the proposed GNN-RSMA algorithm under Algorithm~\ref{alg:gnn}, showing the evolution of the training loss (top panel) together with the resulting $\mathrm{SE}_\mathrm{min}$ (middle panel) and Jain's fairness index (bottom panel). In each panel, the shaded region represents the raw values per-epoch, while the solid line denotes the moving average over a $50$-epoch window, highlighting the underlying trend amid epoch-to-epoch fluctuations. The loss (top panel) declines sharply within the first few thousand iterations before gradually flattening, and the $\textrm{SE}_\textrm{min}$ and Jain's fairness index (middle and bottom panels) improve correspondingly, both saturating close to a stable value within the same early phase and remaining stable thereafter. These results demonstrate that the proposed training procedure converges reliably, reaching a stable operating point well before the full training budget of $100{,}000$ epochs is exhausted.

\subsection{Performance Comparison with SCA Approach}
Fig.~\ref{fig:cdf} compares the cumulative distribution function (CDF) of $\textrm{SE}_\textrm{min}$ over $1000$ random network realizations for the proposed GNN-RSMA algorithm and the SCA-based RSMA, NOMA, and SDMA benchmarks.
The proposed GNN-RSMA scheme tracks the SCA-RSMA benchmark closely and lies well to the right of the NOMA and SDMA baselines across the entire distribution. The gap over the baselines reflects RSMA's greater flexibility in managing inter-cluster interference relative to NOMA and SDMA's fixed decoding strategies, while the small gap to SCA-RSMA indicates that the learned GNN-RSMA policy closely approximates the SCA benchmark allocation. At the median, GNN-RSMA attains a $\textrm{SE}_\textrm{min}$ of $0.493$~b/s/Hz, compared with $0.536$~b/s/Hz for the SCA-RSMA upper bound, $0.302$~b/s/Hz for NOMA, and $0.124$~b/s/Hz for the SDMA case. Consequently, the proposed GNN-RSMA scheme achieves approximately $92\%$ of the SCA performance. Crucially, whereas SCA achieves its suboptimal performance through an iterative optimization procedure with high per-instance computational cost, the proposed GNN attains comparable $\textrm{SE}_\textrm{min}$ with a few low-complexity forward passes, making it well suited for real-time deployment. 

\begin{figure}
    \centering
    \includegraphics[width=0.7\linewidth]{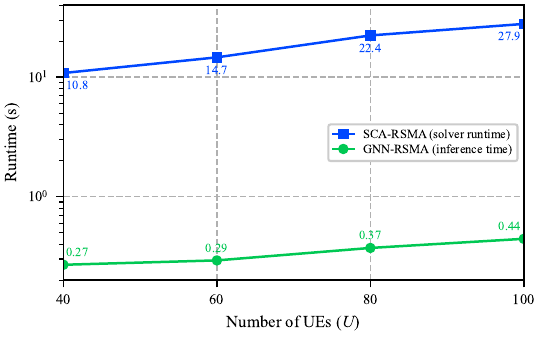}
    \caption{\small Run time vs. number of UEs for the proposed GNN-RSMA (inference time) and the SCA-RSMA (solver runtime).}
    \label{fig:runtime_vs_U}
\end{figure}

Accordingly, Fig.~\ref{fig:runtime_vs_U} compares the run time of the proposed GNN-RSMA algorithm and the conventional SCA-RSMA baseline as a function of the number of UEs, $U$, ranging from $40$ to $100$. The proposed GNN-RSMA algorithm consistently achieves significantly lower run time, requiring only $0.27$--$0.44$ s compared with $10.8$--$27.9$ s for SCA-RSMA. This corresponds to an approximately forty-fold reduction in run time at $U=40$, increasing to more than a sixty-fold reduction at $U=100$, demonstrating the superior scalability of the proposed approach. The improvement stems from replacing the iterative optimization of SCA with a small number of forward passes, which directly infer the RSMA power allocation and common-rate splits while retaining performance close to the benchmark.

\begin{figure}[t]
    \centering
    \includegraphics[width=0.7\linewidth]{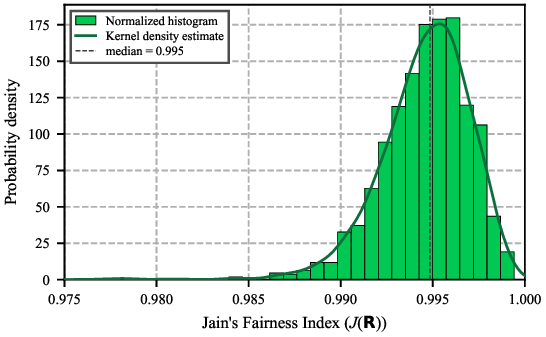}
    \caption{\small Probability distribution of $J(\mathbf{R})$.}
    \label{fig:fairness}
\end{figure}
In addition, Fig.~\ref{fig:fairness} shows the distribution of Jain's fairness index, $J(\mathbf{R})$, over all evaluated network realizations. The distribution is tightly concentrated near unity, with a median of $0.995$, showing that the proposed algorithm achieves excellent fairness among UEs. The narrow histogram and the kernel density estimate confirm that fairness remains stable from one realization to the next. Moreover, almost no realizations fall at lower fairness values, so UE rates stay nearly uniform throughout.

\subsection{Generalization with Respect to the Number of UEs and Impact of WU-Clustering}
\label{subsec:generalization_nu_clustering}

\begin{figure}
    \centering
    \includegraphics[width=0.7\linewidth]{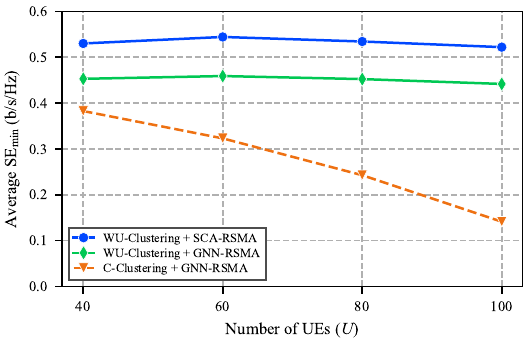}
    \caption{\small Average $\textrm{SE}_\textrm{min}$ vs. number of UEs for GNN-RSMA and SCA-RSMA under WU-Clustering and C-Clustering.}
    \label{fig:minse_vs_nu}
\end{figure}
% Fig.~\ref{fig:minse_vs_nu} plots the average $\textrm{SE}_\textrm{min}$ versus the number of UEs $U$ and compares the performance of GNN-RSMA to that of SCA-RSMA under WU-Clustering. The GNN is trained on variable-size UE sets and evaluated together with the SCA-RSMA for $U \in \{40, 60, 80, 100\}$ with $R = 10$ orthogonal RBs, averaging $1000$ independent random realizations. It can be observed that under WU-Clustering, the performance of GNN-RSMA and SCA-RSMA remains nearly constant. This is noteworthy because for a fixed number of RBs ($R$), increasing the number of UEs ($U$) increases the number of clusters sharing each RB, thereby increasing the co-channel interference experienced by each UE.
% The near-constant $\mathrm{SE}_\mathrm{min}$ therefore indicates that WU-Clustering, together with the MMF--RSMA--PA, compensates for this growing contention, and that the GNN reproduces the same behavior. Although the generalized GNN, trained over variable network sizes, achieves a slightly lower $\textrm{SE}_\textrm{min}$ than a GNN trained exclusively for $U=60$, this modest performance gap is the expected tradeoff for improved generalization across different network sizes.
Fig.~\ref{fig:minse_vs_nu} plots the average $\textrm{SE}_\textrm{min}$ versus the number of UEs $U$ and compares the performance of GNN-RSMA to that of SCA-RSMA under WU-Clustering. The GNN is trained on variable-size UE sets and evaluated together with the SCA-RSMA for $U \in \{40, 60, 80, 100\}$ with $R = 10$ orthogonal RBs, averaging $1000$ independent random realizations. Since $L = \lceil U/R \rceil$, both the number of UEs and the number of clusters change across this sweep, so each setting yields a different input graph. However, owing to the GNN architecture, a single model can be trained and applied across different numbers of UEs, whereas conventional DNN- and DRL-based policies require a separate model for each network size. Furthermore, it can be observed that the performance of both GNN-RSMA and SCA-RSMA remains nearly constant in $U$. This is noteworthy because, for a fixed $R$, increasing $U$ increases the number of clusters sharing each RB, thereby increasing the co-channel interference experienced by each UE. The near-constant $\mathrm{SE}_\mathrm{min}$ therefore indicates that WU-Clustering, together with the MMF--RSMA--PA, compensates for this growing contention. Although the generalized GNN achieves a slightly lower $\mathrm{SE}_\mathrm{min}$ than a GNN trained exclusively for $U=60$, this modest gap is the expected tradeoff for a single policy covering the entire range of network sizes.

To validate that the observed flatness stems from the WU-Clustering scheme, Fig.~\ref{fig:minse_vs_nu} also plots the performance of GNN-RSMA with a centroid-based clustering (C-clustering) baseline, in which each beam is steered toward the centroid of its associated UEs while all other simulation parameters remain unchanged. As shown by the dashed curve, the average $\textrm{SE}_\textrm{min}$ achieved by GNN-RSMA under C-clustering decreases monotonically from $0.383$ b/s/Hz at $U=40$ to 0.141 b/s/Hz at $U=100$, corresponding to a $63\%$ reduction, compared with only about $2\%$ under WU-Clustering. These results demonstrate that the superior scalability of WU-Clustering stems from its worst-UE-aware beam design, which effectively mitigates the increase in co-channel interference as the number of UEs grows.

\subsection{Impact of the GNN Refinement Passes ($N_\mathrm{p}$) and Message Passing Layers ($K$)}
\begin{figure}
    \centering
    \includegraphics[width=0.7\linewidth]{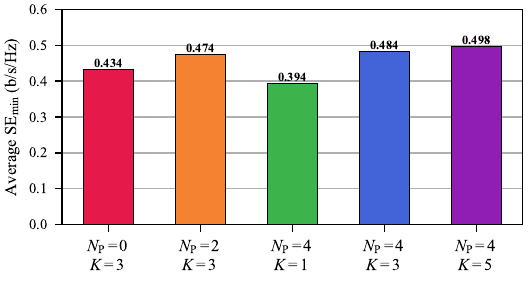}
    \caption{\small Average $\textrm{SE}_\textrm{min}$ vs. number of refinement passes and message-passing layers for the GNN-RSMA.}
    \label{fig:refinement}
\end{figure}
Fig.~\ref{fig:refinement} plots the average $\mathrm{SE}_\mathrm{min}$ as a function of the number of refinement passes $(N_{\mathrm{p}})$ and the number of message-passing layers $(K)$ averaged over $1000$ independent random network realizations.
It can be observed that increasing $N_{\mathrm{p}}$ from $0$ to $2$ improves the average $\textrm{SE}_\textrm{min}$ approximately by $9.2\%$, indicating that even a small number of message-passing refinement iterations allows the GNN to better identify and boost the most disadvantaged UEs in each allocation. Extending the number of refinement passes further to $N_{\mathrm{p}} = 4$ yields an additional increase of about $2.1\%$ over $N_{\mathrm{p}} = 2$. This trend shows that iterative refinement primarily benefits the weakest UEs by allowing information to propagate more effectively across the graph, progressively correcting suboptimal allocations from earlier passes and raising the achieved $\mathrm{SE}_\mathrm{min}$. 
However, the decreasing gains suggest that most of the benefit to the worst-UEs is captured within the first couple of refinement iterations, after which additional passes contribute only marginally while increasing computational cost.

Moreover, Fig.~\ref{fig:refinement} also illustrates the effect of the number of message-passing layers $K$, evaluated at a fixed $N_{\mathrm{p}} = 4$. Increasing the depth from $K = 1$ to $K = 3$ raises the average $\textrm{SE}_\textrm{min}$ by $22.8\%$, since deeper message passing enables each UE's representation to aggregate information from a larger neighborhood, allowing the model to resolve inter-UE interference and protect the weakest UEs more effectively. Extending the depth further to $K = 5$ provides only a marginal additional improvement by $2.9\%$ over $K = 3$, mirroring the diminishing-returns behavior observed for the refinement passes. These observations suggest that a moderate number of message-passing layers and refinement passes is sufficient to capture the relevant interference dependencies while maintaining computational efficiency.

\subsection{Impact of HAPS UPA Architecture}
\begin{figure}
    \centering
    \includegraphics[width=0.7\linewidth]{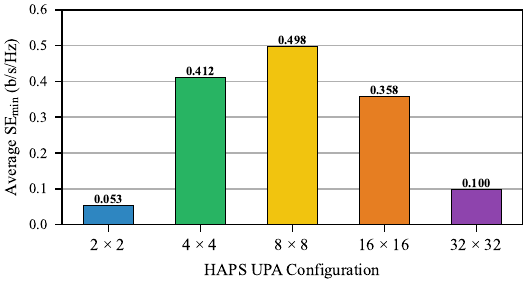}
    \caption{\small Average $\textrm{SE}_\textrm{min}$ vs. number of antennas at HAPS for the GNN-RSMA.}
    \label{fig:HAPSAntenna}
\end{figure}

Fig.~\ref{fig:HAPSAntenna} compares the average $\textrm{SE}_\textrm{min}$ achieved with different HAPS UPA configurations. As the number of antenna elements increases, the beamwidth around the boresight decreases, resulting in higher array gain and improved spatial suppression of co-channel interference. While narrower beams enhance the beamforming gain for UEs located close to the boresight, they reduce the gain for UEs with larger angular offsets from the boresight.
At the same time, narrower beams limit the leakage of interference toward other clusters, thereby improving inter-cluster interference management. This leads to a fundamental trade-off between beamforming gain and interference mitigation.
For the considered network parameters, increasing the antenna size up to $8 \times 8$ improves performance due to more effective interference suppression, yielding the highest average $\textrm{SE}_\textrm{min}$ of $0.498$~b/s/Hz. However, further increasing the number of antenna elements results in performance degradation, as the reduced coverage of narrower beams negatively impacts UEs located away from the beam center, with the $\textrm{SE}_\textrm{min}$ dropping to $0.10$~b/s/Hz at $32 \times 32$.

\section{Conclusion} \label{Sec:Conclusion}
This work addresses the max-min fairness (MMF) power allocation problem in large-scale HAPS networks, where many UEs need to be served with limited radio resources, resulting in severe co-channel interference. To tackle this challenge, we proposed a graph neural network (GNN)-based RSMA framework that models UEs, clusters, and RBs as graph nodes and jointly optimizes common- and private-stream power allocation as well as common-rate sharing through a small number of low-complexity forward passes. The model is trained in a fully unsupervised manner using a smooth MMF surrogate objective and enforces power constraints through a softmax-based decoder. Simulation results show that GNN-RSMA achieves comparable fairness and worst-user performance to that of the iterative SCA benchmark at a fraction of the computational cost, with the runtime advantage widening as network size grows, while generalizing across variable network sizes without retraining. These results establish the proposed GNN-RSMA as a scalable, practical alternative to convex-optimization-based interference management for real-time RSMA deployment.

\bibliographystyle{IEEEtran}
\bibliography{references}

\end{document}